\documentstyle{article}

\newcommand{\be}{\begin{equation}}
\newcommand{\ee}{\end{equation}}
\newcommand{\bea}{\begin{eqnarray}}
\newcommand{\eea}{\end{eqnarray}}
\newcommand{\ep}{\varepsilon}

\begin{document}
\thispagestyle{empty}
 
 
\begin{flushright} 
{\large INLO-PUB--5/96} \\[2mm]
{\large hep-ph/9606244} \\[9mm]
{\large June 1996}           
\end{flushright}

\vspace{30mm}

\begin{center}
{\bf \Large
Small-threshold behaviour of two-loop self-energy diagrams: \\[2mm]
       some special cases${}^{*\dag}$}
\end{center}
\vspace{1cm}
\begin{center}
{\large
F.~A.~Berends$^{a}$, \ \
A.~I.~Davydychev$^{a,b}$ \ \
 \ \ and \ \
V.~A.~Smirnov$^{a,b}$ }
 
\vspace{1cm}
{\large 
$^{a}${\em Instituut-Lorentz, \ \ \ \ University \ of \ Leiden, \\ 
             P.O.B. 9506, 2300 RA Leiden, The Netherlands} }
\\
\vspace{.3cm}
{\large 
$^{b}${\large \em Institute for Nuclear Physics, Moscow State University, \\
           119899 Moscow, Russia} }
\end{center}

\vspace{20mm}

\begin{abstract}
{\large
An algorithm to construct analytic approximations to two-loop diagrams
describing their behaviour at small non-zero thresholds is discussed.
For some special cases (involving two different-scale mass parameters),
several terms of the expansion are obtained.}
\end{abstract}

\vspace{20mm}

\begin{minipage}{14cm}
\hrulefill

${}^*$ Presented at 
{\sl QCD and QED in Higher Order}, 1996 Zeuthen Workshop on 
Elementary Particle Theory, Rheinsberg, April 21--26, 1996,
to appear in Nucl.\ Phys.\ B (Proc. Suppl.). \\[1mm]
${}^{\dag}$ This research was supported by the EU
       under contract number INTAS-93-0744.
\end{minipage}

\newpage

\section{INTRODUCTION}

The calculation of two-loop Feynman diagrams with (different) masses
meets serious technical difficulties, even for two-point functions.
After the tensor decomposition of two-loop self-energy diagrams
\cite{WSB}, the problem is reduced to scalar integrals.
When no exact analytic expressions are available, two main strategies 
can be chosen: either numerical or approximative analytical.  
As to the numerical strategy,
there exist various integral representations \cite{int_rep} 
which provide results for given masses via numerical
integration. A semi-numerical approach was considered in \cite{FT}. 

The analytical approach uses explicit formulae for the
asymptotic expansion of Feynman diagrams in momenta and masses and is
based on general theorems on asymptotic expansions  \cite{as-ex} 
(see also \cite{Smirnov} for review). 
For two-loop self-energy diagrams with general masses, 
the first results have been obtained
outside the particle thresholds: a small momentum expansion
below the lowest threshold \cite{DT1} and a large momentum expansion
above the highest threshold \cite{DST} (some special
cases were considered in \cite{special}). 
However, small and large momentum expansions
do not describe the behaviour between the  
lowest and the highest physical thresholds. In the special
case of a zero mass threshold the small momentum expansion
can be extended to the lowest non-vanishing threshold \cite{BDST}
(see also in \cite{LvRV}).
The expansion coefficients involve then powers of $\ln(-k^2)$
(where $k$ is the external momentum). 

Our investigation considers 
cases when one (or two) two-particle threshold(s) is (are) small 
with respect to the other thresholds, but not anymore zero. Using 
asymptotic expansions  in the large mass limit,
a series converging above the
small threshold can be found. The expansion coefficients
now contain the two-particle cut(s) associated with the small
threshold(s) and therefore the non-regular behaviour around the 
threshold(s) is described. Here we restrict the applications to 
diagrams involving one large ($M$) and one small ($m$) mass parameter, 
i.e.\ some masses are equal.
In \cite{BDS} a more detailed description of the approach
and of the expressions for general masses is given.

\section{THE APPROACH}

Consider the diagram shown in Fig.~1. The corresponding 
scalar Feynman integral reads
\be
\label{defJ}
J\left( \{\nu_i\} ; \{m_i\} ; k \right)
= \int \int 
\frac{\mbox{d}^n p \; \mbox{d}^n q}
{D_1^{\nu_1} D_2^{\nu_2} D_3^{\nu_3} D_4^{\nu_4} D_5^{\nu_5}} ,
\ee
where $n= 4-2\ep$ is the space-time dimension \cite{dimreg}, 
and $(D_i)^{\nu_i} \equiv (p_i^2 - m_i^2 + i0)^{\nu_i}$ are the powers
of the propagators corresponding 
to the internal lines. The momenta $p_i$ are constructed from
$k$ and the loop momenta.

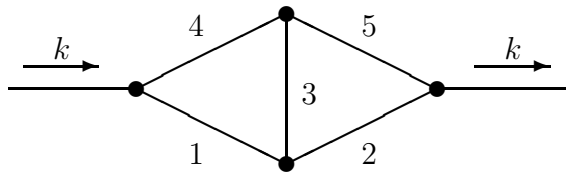
\begin{figure}[b]
\unitlength=1mm
\begin{center}
\begin{picture}(74,22)(0,-11)
\thicklines
\put(17,0){\line(-1,0){17}}
\put(57,0){\line(1,0){17}}
\put(17,0){\line(2,1){20}}
\put(17,0){\line(2,-1){20}}
\put(57,0){\line(-2,1){20}}
\put(57,0){\line(-2,-1){20}}
\put(37,-10){\line(0,1){20}}
\put(17,0){\circle*{2}}
\put(57,0){\circle*{2}}
\put(37,-10){\circle*{2}}
\put(37,10){\circle*{2}}
\put(24,-10){\mbox{\large{1}}}
\put(24,7){\mbox{\large{4}}}
\put(47,-10){\mbox{\large{2}}}
\put(47,7){\mbox{\large{5}}}
\put(39,-2){\mbox{\large{3}}}
\put(2,3){\vector(1,0){10}}
\put(62,3){\vector(1,0){10}}
\put(6,4){\mbox{\large{$k$}}}
\put(66,4){\mbox{\large{$k$}}}
\end{picture}
\end{center}
\caption{The master two-loop two-point diagram}
\end{figure}

In general, the diagram in Fig.~1 has two two-particle thresholds,
at $k^2=(m_1\!+\!m_4)^2$ and $k^2=(m_2\!+\!m_5)^2$, 
and two three-particle thresholds, at $k^2\!=\!(m_1\!+\!m_3\!+\!m_5)^2 \!$
and $k^2\!=\!(m_2\!+\!m_3\!+\!m_4)^2\!$. 
Let us consider some of the masses to be large, 
while the other masses and the 
external momentum are small.
We shall denote the large masses with capital letters, $M_i$. 
The classification of all different small-threshold configurations
has been given in refs.~\cite{BDST,BDS} (see also in Table~1).

We now need to introduce some notation. 
Let $\Gamma$ be the original graph (Fig.~1), its subgraphs 
are denoted as $\gamma$, and the corresponding reduced graph
$\Gamma/\gamma$ is obtained from $\Gamma$ by shrinking 
$\gamma$ to a point. Furthermore, $J_{\gamma}$ is the
di\-men\-sionally-regularized Feynman integral with the denominators 
corresponding to a graph $\gamma$. In particular, $J_{\Gamma}$ 
corresponds to (\ref{defJ}) itself.
Then the general theorem \cite{as-ex} yields
\be
\label{theorem}
J_{\Gamma}  \begin{array}{c} \frac{}{}  \\[1mm]
                    {\mbox{\Huge$\sim$}} \\ {}_{k, \; m_i \to 0}
            \end{array}
\sum_{\gamma} J_{\Gamma / \gamma}
\circ {\cal T}_{k, \; m_{i}, \; q_{i}} \; J_{\gamma} ,
\ee
where the sum goes over all subgraphs $\gamma$ which
(i) contain all the lines with the large masses $M_i$,
and (ii) are one-particle irreducible with respect to the
light lines (with the masses $m_i$).
The Taylor operator ${\cal T}_{k, \; m_{i}, \; q_{i}}$  
expands the integrand of $J_{\gamma}$ in small masses $m_i$, 
the external momentum $k$ and the loop momenta $q_i$ 
which are ``external'' for a given subgraph $\gamma$.
The symbol ``$\circ$'' means that the polynomial in $q_i$,
which appears as a result of applying ${\cal T}$ to $J_{\gamma}$,
should be inserted in the numerator of the integrand
of $J_{\Gamma / \gamma}$. 

Let us consider which subgraphs $\gamma$ contribute to the sum
(\ref{theorem}) for the different small-threshold configurations,
using the numbering in Fig.1.
For example, the ``full'' graph $\Gamma\equiv \{12345\}$ includes
all five lines, $\{134\}$ corresponds to a subgraph without
the lines 2 and 5, etc. The $\gamma$'s contributing
to the expansion are shown in Table~1.

\begin{table}[t]
\caption{Subgraphs $\gamma$ to be included in the expansion.} 
\begin{center}
\begin{tabular}{|l|l|l|} 
\hline
{Case}& {$\!\!\! \begin{array}{c} \mbox{Heavy} \\ \mbox{lines} 
               \end{array} \!\!\! $} 
& {Subgraphs $\gamma$} \\ 
\hline \hline
1  & 1, 3, 4 & $\Gamma,  \{134\} $  \\
1a & 1, 4    & $\Gamma,  \{1245\}, \; \{134\}, \; \{14\} $ \\
1b & 1, 3    & $\Gamma,  \{1235\}, \; \{134\}, \; \{13\} $ \\
2  & 3       & $\Gamma,  \{134\}, \; \{235\}, \; \{3\} $ \\
3  & 1, 5    & $\Gamma,  \{1235\}, \! \{1345\}, \! \{1245\}, \!
                              \{15\} $ \\
4  & 1       & $\Gamma,  \{1245\}, \; \{134\}, \; \{1\} $ \\
\hline
\end{tabular}
\end{center}
\end{table}

After partial fractioning, 
the following types of contributions arise in 
the coefficients of the small-threshold expansion
(in situations with two-part\-icle small thresholds): 
(a) two-loop vacuum diagrams with two (or one) large-mass lines 
    and one (or two) massless lines;
(b) products of a one-loop massive diagram (with small masses and  
    external momentum $k$) and a one-loop massive tadpole;
(c) products of two one-loop massive diagrams
    with external momentum $k$.
For the diagrams containing a small three-particle threshold, 
we also need results for the diagrams obtained
by shrinking the lines corresponding to large masses. 

\begin{table*}[p]
\caption{Terms of the expansion for the cases 1 and 1a}
\label{table2}
\begin{tabular}{l}
\hline 
Case~1: $\hspace*{10mm} J(M,m,M,M,m; k)$ \\
\hline
$M^2   S_0 =  - {\textstyle{1\over2}} \tau + {\textstyle{1\over2}} L_m 
              - {\textstyle{3\over2}} $ 
\\[1mm]
$M^4   S_1 = k^2 \left( - {\textstyle{1\over24}} \tau 
            + {\textstyle{1\over24}} L_m - {\textstyle{1\over16}} \right) 
           + m^2 \left( - {\textstyle{1\over12}} \tau 
           + {\textstyle{1\over6}} L_m - {\textstyle{13\over36}} \right)$ 
\\[1mm]
$M^6   S_2 = (k^2)^2 \left( - {\textstyle{1\over180}} \tau 
           + {\textstyle{1\over180}} L_m - {\textstyle{1\over180}} \right) 
     + k^2 m^2 \left( - {\textstyle{1\over90}} \tau 
                 + {\textstyle{1\over30}} L_m 
                 - {\textstyle{8\over225}} \right) 
     + m^4 \left( - {\textstyle{1\over60}} \tau 
                  + {\textstyle{1\over20}} L_m 
                  - {\textstyle{19\over200}} \right) $
\\[1mm]
$M^8   S_3 = (k^2)^3 \left( - {\textstyle{1\over1120}} \tau 
+ {\textstyle{1\over1120}} L_m - {\textstyle{3\over4480}} \right) 
      + (k^2)^2 m^2 \left( - {\textstyle{1\over560}} \tau 
+ {\textstyle{1\over140}} L_m - {\textstyle{61\over19600}} \right) $
\\[1mm]
$ \hspace*{11mm}      
+ k^2 m^4 \left( - {\textstyle{1\over336}} \tau 
      + {\textstyle{1\over56}} L_m 
      - {\textstyle{29\over1568}} \right)
      + m^6 \left( - {\textstyle{1\over280}} \tau 
+ {\textstyle{1\over70}} L_m - {\textstyle{743\over29400}} \right) $
\\[1mm]
$M^{10}   S_4 = (k^2)^4 \left( - {\textstyle{1\over6300}} \tau 
+ {\textstyle{1\over6300}} L_m - {\textstyle{1\over10500}} \right) 
      + (k^2)^3 m^2 \left( - {\textstyle{1\over3150}} \tau 
+ {\textstyle{1\over630}} L_m + {\textstyle{4\over99225}} \right) $
\\[1mm]
$ \hspace{11mm}
      + (k^2)^2 m^4 \left( - {\textstyle{1\over1800}} \tau 
+ {\textstyle{1\over168}} L_m - {\textstyle{3067\over1058400}} \right) 
      + k^2 m^6 \left( - {\textstyle{1\over1260}} \tau 
+ {\textstyle{1\over126}} L_m - {\textstyle{1333\over158760}} \right) $
\\[1mm]
$ \hspace{11mm}
      + m^8 \left( - {\textstyle{1\over1260}} \tau 
+ {\textstyle{1\over252}} L_m - {\textstyle{2131\over317520}} \right) $
\\[1mm]
$M^{12}   S_5 = (k^2)^5 \left( - {\textstyle{1\over33264}} \tau 
+ {\textstyle{1\over33264}} L_m - {\textstyle{1\over66528}} \right) 
      + (k^2)^4 m^2 \left( - {\textstyle{1\over16632}} \tau 
+ {\textstyle{1\over2772}} L_m + {\textstyle{1091\over7683984}} \right) $
\\[1mm]
$ \hspace*{11mm}      
      + (k^2)^3 m^4 \left( - {\textstyle{1\over9240}} \tau 
+ {\textstyle{53\over27720}} L_m - {\textstyle{6047\over48024900}} \right)
      + (k^2)^2 m^6 \left( - {\textstyle{1\over5940}} \tau 
+ {\textstyle{23\over5940}} L_m - {\textstyle{191071\over82328400}} \right)$ 
\\[1mm]
$ \hspace*{11mm}      
      + k^2 m^8 \left( - {\textstyle{1\over4752}} \tau 
+ {\textstyle{5\over1584}} L_m - {\textstyle{2153\over627264}} \right)
      + m^{10} \left( - {\textstyle{1\over5544}} \tau 
+ {\textstyle{1\over924}} L_m - {\textstyle{22727\over12806640}} \right) $
\\[1mm]
$M^{14}   S_6 = (k^2)^6 \left( - {\textstyle{1\over168168}} \tau 
+ {\textstyle{1\over168168}} L_m - {\textstyle{1\over392392}} \right) 
+ (k^2)^5 m^2 \left( - {\textstyle{1\over84084}} \tau 
+ {\textstyle{1\over12012}} L_m 
+ {\textstyle{126421\over2164322160}} \right) $
\\[1mm]
$ \hspace*{11mm}      
+ (k^2)^4 m^4 \left( - {\textstyle{1\over45864}} \tau 
+ {\textstyle{43\over72072}} L_m 
+ {\textstyle{433253\over2597186592}} \right) 
+ (k^2)^3 m^6 \left( - {\textstyle{1\over28028}} \tau 
+ {\textstyle{313\over180180}} L_m 
- {\textstyle{6920149\over16232416200}} \right)$ 
\\[1mm]
$ \hspace*{11mm}      
+ (k^2)^2 m^8 \left( - {\textstyle{1\over20020}} \tau 
+ {\textstyle{1\over468}} L_m - {\textstyle{63169\over42162120}} \right) 
+ k^2 m^{10} \left( - {\textstyle{1\over18018}} \tau 
+ {\textstyle{1\over858}} L_m - {\textstyle{100453\over77297220}} \right)$ 
\\[1mm]
$ \hspace*{11mm}      
+ m^{12} \left( - {\textstyle{1\over24024}} \tau 
+ {\textstyle{1\over3432}} L_m 
- {\textstyle{288851\over618377760}} \right) $
\\[1mm]
\hline 
Case~1a: $\hspace*{10mm} J(M,m,m,M,m; k)$ \\
\hline
$M^2   S_0 =  - \tau + L_m - 2  $
\\[1mm]
$M^4   S_1 = k^2 \left( - {\textstyle{1\over12}} \tau 
+ {\textstyle{1\over12}} L_m + {\textstyle{1\over12}} \right) 
+ m^2 \left( - \tau L_m - {\textstyle{3\over2}} \tau 
+ L_m^2 + 2 L_m + 2 \zeta_2 - 3 \right) $
\\[1mm]
$M^6   S_2 = (k^2)^2 \left( - {\textstyle{1\over90}} \tau 
+ {\textstyle{1\over90}} L_m + {\textstyle{23\over540}} \right) 
+ k^2 m^2 \left( {\textstyle{1\over9}} \tau 
+ L_m + {\textstyle{35\over18}} \right)$
\\[1mm]
\hspace*{11mm}
$+ m^4 \left( - 4 \tau L_m - {\textstyle{16\over3}} \tau + 6 L_m^2 + 12 L_m 
        + 12 \zeta_2 - {\textstyle{44\over3}} \right) $
\\[1mm]
$M^8   S_3 = (k^2)^3 \left( - {\textstyle{1\over560}} \tau 
+ {\textstyle{1\over560}} L_m + {\textstyle{17\over1120}} \right)
 + (k^2)^2 m^2 \left( {\textstyle{1\over120}} \tau 
+ {\textstyle{7\over15}} L_m + {\textstyle{91\over90}} \right) $
\\[1mm]
\hspace*{11mm}
$+ k^2 m^4 \!\left( {\textstyle{1\over2}} \tau L_m 
\!+\! {\textstyle{29\over24}} \tau \!+\! {\textstyle{3\over2}} L_m^2 
\!+\! {\textstyle{25\over2}} L_m 
            \!+\! 3 \zeta_2 \!+\! {\textstyle{99\over8}} \right) 
\!+\! m^6 \!\left( - 15 \tau L_m \!-\! {\textstyle{79\over4}} \tau 
\!+\! 30 L_m^2 \!+\! 64 L_m 
\!+\! 60 \zeta_2 \!-\! {\textstyle{133\over2}} \right) \hspace*{-1mm}$
\\[1mm]
$M^{10}   S_4 = (k^2)^4 \left( - {\textstyle{1\over3150}} \tau 
+ {\textstyle{1\over3150}} L_m + {\textstyle{367\over63000}} \right) 
+ (k^2)^3 m^2 \left( {\textstyle{1\over1050}} \tau 
+ {\textstyle{11\over42}} L_m + {\textstyle{2977\over4200}} \right) $
\\[1mm]
\hspace*{11mm}
$+ (k^2)^2 m^4 \!\left( - {\textstyle{1\over25}} \tau \!+\! 2 L_m^2 
\!+\! {\textstyle{69\over5}} L_m \!+\! 4 \zeta_2 
\!+\! {\textstyle{3064\over225}} \right) 
\!+\! k^2 m^6 \!\left( 4 \tau L_m \!+\! {\textstyle{112\over15}} \tau 
\!+\! 20 L_m^2 \!+\! 110 L_m 
\!+\! 40 \zeta_2 \!+\! {\textstyle{2756\over45}} \right) \hspace{-1mm}$
\\[1mm]
\hspace*{11mm}
$+ m^8 \left( - 56 \tau L_m - {\textstyle{1108\over15}} \tau + 140 L_m^2 
+ {\textstyle{940\over3}} L_m 
                + 280 \zeta_2 - {\textstyle{13058\over45}} \right) $
\\[1mm]
$M^{12}   S_5 = (k^2)^5 \left( - {\textstyle{1\over16632}} \tau 
+ {\textstyle{1\over16632}} L_m + {\textstyle{2531\over997920}} \right) 
+ (k^2)^4 m^2 \left( {\textstyle{1\over7560}} \tau 
+ {\textstyle{209\over1260}} L_m + {\textstyle{39887\over75600}} \right) $
\\[1mm]
\hspace*{11mm}
$+ (k^2)^3 m^4 \left( - {\textstyle{1\over420}} \tau + 2 L_m^2 
+ {\textstyle{2147\over140}} L_m + 4 \zeta_2 
                + {\textstyle{98921\over5040}} \right) $
\\[1mm]
\hspace*{11mm}
$+ (k^2)^2 m^6 \left( - {\textstyle{1\over3}} \tau L_m 
- {\textstyle{29\over30}} \tau + {\textstyle{115\over3}} L_m^2 
                + {\textstyle{3649\over18}} L_m 
+ {\textstyle{230\over3}} \zeta_2 + {\textstyle{4989\over40}} \right) $
\\[1mm]
\hspace*{11mm}
$+ k^2 m^8 \left( {\textstyle{70\over3}} \tau L_m 
+ {\textstyle{118\over3}} \tau + 175 L_m^2 + 800 L_m 
                + 350 \zeta_2 + {\textstyle{9139\over36}} \right) $
\\[1mm]
\hspace*{11mm}
$+ m^{10} \left( - 210 \tau L_m - {\textstyle{1669\over6}} \tau 
+ 630 L_m^2 + 1459 L_m 
                + 1260 \zeta_2 - {\textstyle{4967\over4}} \right) $
\\[1mm]
$M^{14}   S_6 = (k^2)^6 \left( - {\textstyle{1\over84084}} \tau 
+ {\textstyle{1\over84084}} L_m + {\textstyle{4903\over3923920}} \right) 
+ (k^2)^5 m^2 \left( {\textstyle{1\over48510}} \tau
+ {\textstyle{113\over990}} L_m + {\textstyle{395999\over970200}} \right) $
\\[1mm]
\hspace*{11mm}
$+ (k^2)^4 m^4 \left( - {\textstyle{1\over4410}} \tau + 2 L_m^2 
+ {\textstyle{1754\over105}} L_m + 4 \zeta_2 
                 + {\textstyle{221993\over8820}} \right)$
\\[1mm]
\hspace*{11mm} 
$+ (k^2)^3 m^6 \left( {\textstyle{1\over49}} \tau + 56 L_m^2 
+ {\textstyle{1687\over5}} L_m + 112 \zeta_2 
                 + {\textstyle{13072747\over44100}} \right) $
\\[1mm]
\hspace*{11mm}
$+ (k^2)^2 m^8 \left( - 4 \tau L_m - {\textstyle{928\over105}} \tau 
+ 462 L_m^2 + {\textstyle{10794\over5}} L_m 
                 + 924 \zeta_2 + {\textstyle{2837671\over3150}} \right)$
\\[1mm]
\hspace*{11mm}
$+ k^2 m^{10} \left( 120 \tau L_m + {\textstyle{4028\over21}} \tau 
+ 1260 L_m^2 + 5180 L_m 
                 + 2520 \zeta_2 + {\textstyle{90217\over105}} \right)$
\\[1mm]
\hspace*{11mm}
$+ m^{12} \left( - 792 \tau L_m - {\textstyle{36874\over35}} \tau 
+ 2772 L_m^2 + {\textstyle{32914\over5}} L_m 
                 + 5544 \zeta_2 - {\textstyle{5515733\over1050}} \right) $
\\[1mm]
\hline 
\end{tabular}
\end{table*}

\begin{table*}[tb]
\caption{Terms of the expansion for the case 1b}
\label{table3}
\begin{tabular}{l}
\hline
Case~1b : $\hspace*{10mm} J(M,m,M,m,m; k)$ \\
\hline
$M^2   S_0 =  - \tau + L_m + \zeta_2 - 4 $
\\[1mm]
$M^4   S_1 = k^2 \left( - {\textstyle{1\over4}} \tau 
+ {\textstyle{1\over4}} L_m + {\textstyle{1\over2}} \zeta_2 
- {\textstyle{13\over12}} \right) 
+ m^2 \left( - \tau L_m - {\textstyle{4\over3}} \tau + L_m^2 
+ {\textstyle{2\over3}} L_m + 3 \zeta_2 
        - {\textstyle{107\over18}} \right) $
\\[1mm]
$M^6   S_2 = (k^2)^2 \left( - {\textstyle{1\over9}} \tau 
+ {\textstyle{1\over9}} L_m + {\textstyle{1\over3}} \zeta_2 
- {\textstyle{673\over1080}} \right) 
+ k^2 m^2 \left( - \tau L_m - {\textstyle{77\over36}} \tau 
+ L_m^2 + {\textstyle{19\over10}} L_m 
                + 4 \zeta_2 - {\textstyle{12973\over1800}} \right)$ 
\\[1mm]
\hspace*{11mm}
$+ m^4 \left( - 3 \tau L_m - {\textstyle{179\over60}} \tau + 4 L_m^2 
+ {\textstyle{217\over60}} L_m 
                + 9 \zeta_2 - {\textstyle{3298\over225}} \right) $
\\[1mm]
$M^8   S_3 = (k^2)^3 \left( - {\textstyle{1\over16}} \tau 
+ {\textstyle{1\over16}} L_m \!+ {\textstyle{1\over4}} \zeta_2 
\!- {\textstyle{1487\over3360}} \right) 
+ (k^2)^2 m^2 \left( - \tau L_m \!- {\textstyle{329\over120}} \tau 
+ L_m^2 \!+ {\textstyle{183\over70}} L_m\! 
               + 5 \zeta_2 \!- {\textstyle{765467\over88200}} \right) $
\\[1mm]
\hspace*{11mm}
$+ k^2 m^4 \left( - {\textstyle{11\over2}} \tau L_m 
- {\textstyle{47\over6}} \tau + {\textstyle{15\over2}} L_m^2 
+ {\textstyle{5953\over420}} L_m 
               + {\textstyle{39\over2}} \zeta_2 
- {\textstyle{1549283\over58800}} \right) $
\\[1mm]
\hspace*{11mm}
$+ m^6 \left( - 8 \tau L_m - {\textstyle{3317\over420}} \tau 
+ 14 L_m^2 + {\textstyle{1535\over84}} L_m 
               + 29 \zeta_2 - {\textstyle{73547\over1764}} \right) $
\\[1mm]
$M^{10}   S_4 = (k^2)^4 \left( - {\textstyle{1\over25}} \tau 
+ {\textstyle{1\over25}} L_m + {\textstyle{1\over5}} \zeta_2 
                          - {\textstyle{86939\over252000}} \right) $
\\[1mm]
\hspace*{11mm}
$+ (k^2)^3 m^2 \left( - \tau L_m - {\textstyle{241\over75}} \tau 
+ L_m^2 + {\textstyle{1579\over504}} L_m 
                + 6 \zeta_2 - {\textstyle{32459123\over3175200}} \right) $
\\[1mm]
\hspace*{11mm}
$+ (k^2)^2 m^4 \left( - 9 \tau L_m - {\textstyle{8611\over525}} \tau 
+ 12 L_m^2  + {\textstyle{74681\over2520}} L_m 
+ 36 \zeta_2 - {\textstyle{17514967\over396900}} \right) $
\\[1mm]
\hspace*{11mm}
$+ k^2 m^6 \left( - 20 \tau L_m - {\textstyle{20093\over840}} \tau 
+ 41 L_m^2  + {\textstyle{117043\over1260}} L_m + 90 \zeta_2 
- {\textstyle{138737377\over1587600}} \right) $
\\[1mm]
\hspace*{11mm}
$+ m^8 \left( - 22 \tau L_m - {\textstyle{8231\over360}} \tau 
+ 49 L_m^2 + {\textstyle{39869\over504}} L_m 
            + 99 \zeta_2 - {\textstyle{41100499\over317520}} \right) $
\\[1mm]
$M^{12}   S_5 = 
(k^2)^5 \left( - {\textstyle{1\over36}} \tau 
+ {\textstyle{1\over36}} L_m + {\textstyle{1\over6}} \zeta_2 
- {\textstyle{2828267\over9979200}} \right) $
\\[1mm]
\hspace*{11mm}
$+ (k^2)^4 m^2 \left( - \tau L_m - {\textstyle{2267\over630}} \tau 
+ L_m^2 + {\textstyle{49103\over13860}} L_m 
 + 7 \zeta_2 - {\textstyle{4534026869\over384199200}} \right) $
\\[1mm]
\hspace*{11mm}
$+ (k^2)^3 m^4 \left( - {\textstyle{27\over2}} \tau L_m 
- {\textstyle{16409\over560}} \tau + {\textstyle{35\over2}} L_m^2 
                + {\textstyle{18179\over360}} L_m 
+ 60 \zeta_2 
- {\textstyle{20105059\over285120}} \right) $ 
\\[1mm]
\hspace*{11mm}
$+ (k^2)^2 m^6 \left( - {\textstyle{136\over3}} \tau L_m 
- {\textstyle{62402\over945}} \tau + {\textstyle{283\over3}} L_m^2 
                + {\textstyle{2806913\over10395}} L_m + 222 \zeta_2 
                - {\textstyle{21361520641\over144074700}} \right) $
\\[1mm]
\hspace*{11mm}
$+ k^2 m^8 \left( - {\textstyle{379\over6}} \tau L_m 
- {\textstyle{1069639\over15120}} \tau + 206 L_m^2 
            + {\textstyle{9873841\over18480}} L_m 
+ {\textstyle{849\over2}} \zeta_2 
- {\textstyle{6628516487\over21344400}} \right) $
\\[1mm]
\hspace*{11mm}
$+ m^{10} \left( - 64 \tau L_m - {\textstyle{1943093\over27720}} \tau 
+ 175 L_m^2  + {\textstyle{8884391\over27720}} L_m + 351 \zeta_2 
- {\textstyle{81848153249\over192099600}} \right) $
\\[1mm]
$M^{14}   S_6 = 
(k^2)^6 \left( - {\textstyle{1\over49}} \tau + {\textstyle{1\over49}} L_m 
+ {\textstyle{1\over7}} \zeta_2 
- {\textstyle{85046849\over353152800}} \right)$
\\[1mm]
\hspace*{11mm}
$+ (k^2)^5 m^2 \left( - \tau L_m - {\textstyle{7689\over1960}} \tau + L_m^2 
                + {\textstyle{49963\over12870}} L_m + 8 \zeta_2 
- {\textstyle{435006955477\over32464832400}} \right) $
\\[1mm]
\hspace*{11mm}
$+ (k^2)^4 m^4 \left( - 19 \tau L_m - {\textstyle{829189\over17640}} \tau 
+ 24 L_m^2  + {\textstyle{27760549\over360360}} L_m + 93 \zeta_2 
                - {\textstyle{7029366464489\over64929664800}} \right) $
\\[1mm]
\hspace*{11mm}
$+ (k^2)^3 m^6 \left( - 92 \tau L_m - {\textstyle{557275\over3528}} \tau 
+ 186 L_m^2 + {\textstyle{221210659\over360360}} L_m + 472 \zeta_2 
                - {\textstyle{14327879597579\over64929664800}} \right) $
\\[1mm]
\hspace*{11mm}
$+ (k^2)^2 m^8 \left( - 176 \tau L_m - {\textstyle{6228527\over27720}} \tau 
+ 650 L_m^2 + {\textstyle{756802831\over360360}} L_m + 1375 \zeta_2 
                - {\textstyle{5163216185287\over12985932960}} \right) $
\\[1mm]
\hspace*{11mm}
$+ k^2 m^{10} \left( - 192 \tau L_m - {\textstyle{2955811\over13860}} \tau 
+ 1008 L_m^2 + {\textstyle{4623547\over1638}} L_m + 2034 \zeta_2 
               - {\textstyle{3647519729639\over2951348400}} \right) $
\\[1mm]
\hspace*{11mm}
$+ m^{12} \left( - 196 \tau L_m - {\textstyle{2239113\over10010}} \tau 
+ 637 L_m^2 + {\textstyle{6324508\over5005}} L_m + 1275 \zeta_2 
           - {\textstyle{6314510897209\over4328644320}} \right) $
\\[1mm]
\hline 
\end{tabular}
\end{table*}

\begin{table*}[tb]
\caption{Terms of the expansion for the case 2}
\label{table4}
\begin{tabular}{l}
\hline
Case~2 : $\hspace*{10mm} J(m,m,M,m,m; k)$ \\
\hline
$M^2   S_0 =  - \tau^2 + 2 \tau L_m - 2 \tau
 - L_m^2 + 2 L_m - 2 \zeta_2  $
\\[1mm]
$M^4   S_1 = 
k^2 \left( {\textstyle{1\over2}} \tau^2 - \tau L_m 
+ {\textstyle{1\over2}} \tau + {\textstyle{1\over2}} L_m^2 
- {\textstyle{1\over2}} L_m 
      + \zeta_2 - {\textstyle{3\over2}} \right) 
+ m^2 \left( - 2 \tau^2 + 6 \tau L_m - \tau - 4 L_m^2 
        - 8 \zeta_2 + 10 \right) $
\\[1mm]
$M^6   S_2 = 
(k^2)^2 \left( - {\textstyle{1\over3}} \tau^2 
+ {\textstyle{2\over3}} \tau L_m - {\textstyle{2\over9}} \tau 
- {\textstyle{1\over3}} L_m^2 
          + {\textstyle{2\over9}} L_m 
- {\textstyle{2\over3}} \zeta_2 + 1 \right)$ 
\\[1mm]
\hspace*{11mm}
$+ k^2 m^2 \left( {\textstyle{8\over3}} \tau^2 
- {\textstyle{20\over3}} \tau L_m + {\textstyle{8\over9}} \tau 
           + 4 L_m^2 - 2 L_m + 8 \zeta_2 - 14 \right) $
\\[1mm]
\hspace*{11mm}
$+ m^4 \left( - {\textstyle{16\over3}} \tau^2 + 20 \tau L_m 
+ {\textstyle{10\over3}} \tau - 18 L_m^2 
           - 14 L_m - 36 \zeta_2 + 47 \right) $
\\[1mm]
$M^8   S_3 = 
(k^2)^3 \left( {\textstyle{1\over4}} \tau^2 
- {\textstyle{1\over2}} \tau L_m + {\textstyle{1\over8}} \tau 
+ {\textstyle{1\over4}} L_m^2 
          - {\textstyle{1\over8}} L_m 
+ {\textstyle{1\over2}} \zeta_2 - {\textstyle{115\over144}} \right)$ 
\\[1mm]
\hspace*{11mm}
$+ (k^2)^2 m^2 \left( - 3 \tau^2 + 7 \tau L_m 
- {\textstyle{3\over4}} \tau - 4 L_m^2 
                + {\textstyle{2\over3}} L_m - 8 \zeta_2 
+ {\textstyle{217\over18}} \right) $
\\[1mm]
\hspace*{11mm}
$+ k^2 m^4 \left( 12 \tau^2 - 35 \tau L_m - {\textstyle{11\over4}} \tau 
+ 22 L_m^2 - {\textstyle{34\over3}} L_m + 44 \zeta_2 
- {\textstyle{3185\over36}} \right) $
\\[1mm]
\hspace*{11mm}
$+ m^6 \left( - 16 \tau^2 + 70 \tau L_m + {\textstyle{149\over6}} \tau
        - 80 L_m^2 - {\textstyle{292\over3}} L_m - 160 \zeta_2 
+ {\textstyle{1819\over9}} \right)$
\\[1mm]
$M^{10}   S_4 = 
(k^2)^4 \left( - {\textstyle{1\over5}} \tau^2 
+ {\textstyle{2\over5}} \tau L_m - {\textstyle{2\over25}} \tau 
          - {\textstyle{1\over5}} L_m^2 + {\textstyle{2\over25}} L_m 
- {\textstyle{2\over5}} \zeta_2 + {\textstyle{23\over36}} \right) $
\\[1mm]
\hspace*{11mm}
$+ (k^2)^3 m^2 \left( {\textstyle{16\over5}} \tau^2 
- {\textstyle{36\over5}} \tau L_m + {\textstyle{16\over25}} \tau 
               + 4 L_m^2 - L_m + 8 \zeta_2 - {\textstyle{245\over18}} 
\right)$ 
\\[1mm]
\hspace*{11mm}
$+ (k^2)^2 m^4 \left( - {\textstyle{96\over5}} \tau^2 
+ {\textstyle{252\over5}} \tau L_m + {\textstyle{58\over25}} \tau 
                - 34 L_m^2 - {\textstyle{47\over3}} L_m - 68 \zeta_2 
+ {\textstyle{3187\over36}} \right) $
\\[1mm]
\hspace*{11mm}
$+ k^2 m^6 \left( {\textstyle{256\over5}} \tau^2 - 168 \tau L_m 
- {\textstyle{556\over15}} \tau + 100 L_m^2 
            - {\textstyle{274\over3}} L_m + 200 \zeta_2 
- {\textstyle{8509\over18}} \right) $
\\[1mm]
\hspace*{11mm}
$+ m^8 \left( - {\textstyle{256\over5}} \tau^2 + 252 \tau L_m 
+ {\textstyle{603\over5}} \tau - 350 L_m^2 
        - {\textstyle{1573\over3}} L_m - 700 \zeta_2 
+ {\textstyle{30475\over36}} \right) $
\\[1mm]
$M^{12}   S_5 = 
(k^2)^5 \left( {\textstyle{1\over6}} \tau^2 
- {\textstyle{1\over3}} \tau L_m + {\textstyle{1\over18}} \tau 
+ {\textstyle{1\over6}} L_m^2 
          - {\textstyle{1\over18}} L_m + {\textstyle{1\over3}} \zeta_2 
- {\textstyle{973\over1800}} \right)$
\\[1mm]
\hspace*{11mm}
$+ (k^2)^4 m^2 \left( - {\textstyle{10\over3}} \tau^2 
+ {\textstyle{22\over3}} \tau L_m - {\textstyle{5\over9}} \tau 
                - 4 L_m^2 + {\textstyle{8\over15}} L_m 
- 8 \zeta_2 + {\textstyle{944\over75}} \right) $
\\[1mm]
\hspace*{11mm}
$+ (k^2)^3 m^4 \left( {\textstyle{80\over3}} \tau^2 
- 66 \tau L_m - 2 \tau + 38 L_m^2 
                - {\textstyle{449\over30}} L_m + 76 \zeta_2 
- {\textstyle{34454\over225}} \right) $
\\[1mm]
\hspace*{11mm}
$+ (k^2)^2 m^6 \left( - {\textstyle{320\over3}} \tau^2 + 308 \tau L_m 
+ {\textstyle{440\over9}} \tau 
                - {\textstyle{772\over3}} L_m^2 
- {\textstyle{4474\over15}} L_m - {\textstyle{1544\over3}} \zeta_2 
                + {\textstyle{708707\over1350}} \right) $
\\[1mm]
\hspace*{11mm}
$+ k^2 m^8 \left( {\textstyle{640\over3}} \tau^2 - 770 \tau L_m 
- {\textstyle{4495\over18}} \tau 
            + 385 L_m^2 - {\textstyle{1457\over2}} L_m 
+ 770 \zeta_2 - {\textstyle{827323\over360}} \right) $
\\[1mm]
\hspace*{11mm}
$+ m^{10} \left( - {\textstyle{512\over3}} \tau^2 + 924 \tau L_m 
+ {\textstyle{7867\over15}} \tau 
           - 1512 L_m^2 - {\textstyle{12804\over5}} L_m - 3024 \zeta_2 
+ {\textstyle{87669\over25}} \right) $
\\[1mm]
$M^{14}   S_6 = 
(k^2)^6 \left( - {\textstyle{1\over7}} \tau^2 
+ {\textstyle{2\over7}} \tau L_m - {\textstyle{2\over49}} \tau 
- {\textstyle{1\over7}} L_m^2 
          + {\textstyle{2\over49}} L_m - {\textstyle{2\over7}} \zeta_2 
+ {\textstyle{139\over300}} \right) $
\\[1mm]
\hspace*{11mm}
$+ (k^2)^5 m^2 \left( {\textstyle{24\over7}} \tau^2 
- {\textstyle{52\over7}} \tau L_m + {\textstyle{24\over49}} \tau 
               + 4 L_m^2 - {\textstyle{2\over3}} L_m + 8 \zeta_2 
- {\textstyle{336\over25}} \right) $
\\[1mm]
\hspace*{11mm}
$+ (k^2)^4 m^4 \left( - {\textstyle{240\over7}} \tau^2 
+ {\textstyle{572\over7}} \tau L_m + {\textstyle{86\over49}} \tau 
                - 50 L_m^2 - {\textstyle{272\over15}} L_m 
- 100 \zeta_2 + {\textstyle{116951\over900}} \right) $
\\[1mm]
\hspace*{11mm}
$+ (k^2)^3 m^6 \left( {\textstyle{1280\over7}} \tau^2 
- {\textstyle{3432\over7}} \tau L_m - {\textstyle{8884\over147}} \tau 
                + 256 L_m^2 - {\textstyle{5014\over15}} L_m 
+ 512 \zeta_2 - {\textstyle{106966\over75}} \right) $
\\[1mm]
\hspace*{11mm}
$+ (k^2)^2 m^8 \left( - {\textstyle{3840\over7}} \tau^2 + 1716 \tau L_m 
+ {\textstyle{8873\over21}} \tau
                - 1862 L_m^2 - {\textstyle{50537\over15}} L_m - 3724 \zeta_2 
                + {\textstyle{835633\over300}} \right) $
\\[1mm]
\hspace*{11mm}
$+ k^2 m^{10} \left( {\textstyle{6144\over7}} \tau^2 
- 3432 \tau L_m - {\textstyle{145682\over105}} \tau 
              + 1176 L_m^2 - {\textstyle{78584\over15}} L_m + 2352 \zeta_2 
              - {\textstyle{2356388\over225}} \right) $
\\[1mm]
\hspace*{11mm}
$+ m^{12} \left( - {\textstyle{4096\over7}} \tau^2 + 3432 \tau L_m 
+ {\textstyle{228982\over105}} \tau 
     - 6468 L_m^2 - {\textstyle{178258\over15}} L_m - 12936 \zeta_2 
     + {\textstyle{6500323\over450}} \right) $
\\[1mm]
\hline
\end{tabular}
\end{table*}

The contributions of type (a) are discussed in 
\cite{BDST}. The one-loop two-point functions
involving small masses of internal lines (occurring in the
contributions (b) and (c)) 
are ``responsible'' for describing the
two-particle threshold irregularities (some relevant results
are collected in \cite{BDS}).

\section{SOME RESULTS}

As an example, consider
the diagram in Fig.~1 with 
$\nu_1\!=\ldots =\! \nu_5 \!=\! 1$
and $n\!=\!4$.
Denote the corresponding integral (\ref{defJ}) as
(here, $m_i$ may correspond to either small or large masses)
\be
\label{J-exp}
J(m_1,m_2,m_3,m_4,m_5; k)
= - \pi^4 \sum_{j=0}^{\infty} S_j \; , 
\ee
where $S_j$ are the terms of our expansion.
For a given $j$, the term $S_j$ is a sum of all contributions
of the order $(k^2)^{j_0} \prod (m_i^2)^{j_i}$ (where the product
is taken over all lines with {\em small} masses) with
$j_0+ \sum j_i = j$. 
The $S_j$ may contain 
logarithms of the ratios of the masses and some functions 
which are described below.

The contributions of type (a) involve 
a function ${\cal{H}}(M_1^2, M_2^2)$ 
which can be expressed in terms of $\mbox{Li}_2$ (see in \cite{BDST}).
It is  antisymmetric and therefore vanishes when $M_1^2=M_2^2$.

The contributions of types (b) and (c) are expressed through the
function $\tau(m_1, m_2; k^2)$ which is related to the finite part 
of the one-loop self energy and can be expressed in terms
of elementary functions (see e.g. in \cite{'tHV'79}).
By using the notation
\be
\label{Delta}
\Delta \equiv 4 m_1^2 m_2^2 - (k^2-m_1^2-m_2^2)^2
\ee
we get
\bea
\label{tau}
\tau(m_1, m_2; k^2)  
= \frac{1}{2 k^2} \biggl\{ \sqrt{-\Delta} \;
\ln\frac{k^2-m_1^2-m_2^2-\sqrt{-\Delta}}
        {k^2-m_1^2-m_2^2+\sqrt{-\Delta}}
+ (m_1^2 - m_2^2) \ln\frac{m_2^2}{m_1^2}
\nonumber \\
+ \mbox{i} \pi \sqrt{-\Delta} \; 
\theta\left(k^2-(m_1+m_2)^2\right) \biggr\} .
\eea
In our contributions, the $\tau$ function depends on the small masses
only. In fact, it contains the main information
about the small-threshold behaviour at two-particle thresholds.
The $\theta$ term in the braces yields an imaginary part in the
region beyond the physical threshold (i.e. for $k^2 > (m_1+m_2)^2$).
Further properties of the $\tau$ function are discussed in \cite{BDS}.

In Tables~2, 3, 4 
we present results for the terms of 
the small-threshold expansion (\ref{J-exp}) for the cases~1, 1a, 1b and 2
(cf. Table~1), provided that all large masses are equal to $M$
and all small masses are equal to $m$. 
We use the notation
\[
\tau \equiv \tau(m,m; k^2), \hspace{3mm}
L_m \equiv \ln(m^2/M^2), \hspace{3mm}
\zeta_2 = {\textstyle{1\over6}} \pi^2 . \hspace*{-3mm}
\]
In the limit $m\to 0$, $\tau \Rightarrow (L_m - L)$,
where $L \equiv \ln(-k^2/M^2)$. 
In this limit the results presented in Tables~2--4
reproduce the first four columns of Table~1 from ref.~\cite{BDST}.
To perform the calculations, we used the {\sf REDUCE} system 
\cite{Reduce}. 

\section{CONCLUSIONS}

We have studied the behaviour of 
two-loop self-energy diagrams when the external momentum 
and some of the masses are small with respect to the large masses.
By use of explicit formulae for the terms of asymptotic expansions 
in the large mass limit, we presented an analytic approach to
calculating these diagrams by keeping the first few terms 
of the expansion. The main idea was to avoid putting
any conditions on relative values of the external momentum squared
and small masses.
In addition to the general results presented in \cite{BDS}, here
we listed the contributions (from $S_0$ to $S_6$) for some 
special cases. Some of the diagrams considered are interesting from
the physical point of view, since they may occur in the Standard Model
calculations.   

In cases~3 and 4, three-particle small thresholds arise. 
The small-threshold behaviour 
is then defined by the functions corresponding to the sunset
diagram (with three propagators) and the diagram with four propagators
(e.g. eq.~(\ref{defJ}) with $\nu_1=0$).
Unfortunately, sufficient {\em analytic} information about 
these diagrams is not yet available.

The algorithm can be also extended to the
three-point two-loop diagrams with small thresholds (for zero-mass 
thresholds, see ref.~\cite{FST}).

\vspace{3mm}

{\bf Acknowledgements.}
A.~D. is grateful to the organizers (DESY-Zeuthen) for support
of his participation in the Rheinsberg workshop. 
A.~D. and V.~S. were partly supported by the RFBR grant 96-01-00654.


\end{document}